\documentclass{jfm}

\usepackage[english]{babel}
\usepackage[T1]{fontenc}
\usepackage{graphicx}

\usepackage{newtxtext}
\usepackage{newtxmath}
\usepackage{comment}
\usepackage{amsfonts,mathtools,amssymb,stmaryrd,dsfont}
\usepackage{color,xcolor}
\usepackage[breaklinks,colorlinks=true,allcolors=blue]{hyperref}
\usepackage{epstopdf, epsfig}
\usepackage{subfig}
\usepackage{natbib}
\usepackage{grffile}
\usepackage{amsmath}
\usepackage{extarrows}

\definecolor{ForestGreen}{RGB}{34,139,34}
\definecolor{NavyBlue}{RGB}{0,0,128}

\newcommand{\bX}{{\ensuremath\boldsymbol{X}}}

\newcommand{\bx}{{\ensuremath\boldsymbol{x}}}
\newcommand{\bee}{{\ensuremath\boldsymbol{e}}}

\newcommand{\bu}{{\ensuremath\boldsymbol{u}}}
\newcommand{\bv}{{\boldsymbol{\text{\bf v}}}}
\newcommand{\bU}{{\ensuremath\boldsymbol{U}}}
\newcommand{\bV}{{\ensuremath\boldsymbol{V}}}
\newcommand{\bR}{{\ensuremath\boldsymbol{R}}}
\newcommand{\bS}{{\ensuremath\boldsymbol{S}}}

\newcommand{\bs}{{\ensuremath\boldsymbol{s}}}
\newcommand{\bff}{{\ensuremath\boldsymbol{f}}}
\newcommand{\bfF}{{\ensuremath\boldsymbol{F}}}

\newcommand{\bk}{{\ensuremath\boldsymbol{k}}}

\newcommand{\be}{\begin{equation}}
\newcommand{\ee}{\end{equation}}
\newcommand{\bea}{\begin{eqnarray}}
\newcommand{\eea}{\end{eqnarray}}

\newcommand{\av}[1]{\left\langle{#1}\right\rangle}
\newcommand{\fil}[2]{#1_{#2}}

\newcommand{\impa}{Instituto de Matemática Pura e Aplicada -- IMPA, Rio de Janeiro, Brazil}
\newcommand{\inphyni}{Universit\'e C\^ote d'Azur, CNRS, Institut de Physique de Nice, 06200 Nice, France}
\newcommand{\UTV}{University of Rome " Tor Vergata", Department of Physics and INFN, Rome, Italy}
\title{Scale invariance of intermittency in LES  turbulence}

\author{B. Magacho\aff{1\corresp{\email{bruno.magacho@impa.br}}}, S. Thalabard\aff{2}, M. Buzzicotti\aff{3}, F. Bonaccorso\aff{3}, L. Biferale\aff{3}, A. A. Mailybaev\aff{1}}
\affiliation{\aff{1} \impa\\ \aff{2} \inphyni\\ \aff{3} \UTV} 
\date{\today}

\newcommand\corr{\color{black}}
\newcommand\rroc{\color{black}}
\usepackage{titlesec}
\titleformat{\subsubsection}[runin]{\large \itshape}{}{}{}[.\;]
\graphicspath{{./figs/}{./}}
\begin{document}
\maketitle

\begin{abstract}
Turbulent flows exhibit large intermittent fluctuations from inertial to dissipative scales,  characterized by multifractal statistics and breaking the statistical self-similarity.  It has recently been proposed that the Navier-Stokes turbulence restores a hidden form of scale invariance in the inertial interval when formulated for a dynamically (nonlinearly) rescaled quasi-Lagrangian velocity field. Here we show that such hidden self-similarity extends to the Large-Eddy Simulation (LES) approach in computational fluid dynamics (CFD). In particular, we show that classical subgrid-scale models, such as implicit or explicit Smagorinsky closures, respect the hidden scale invariance at all scales -- both resolved and subgrid. 
In the inertial range, they reproduce the hidden scale invariance of  Navier-Stokes statistics. 
These properties are verified very accurately by numerical simulations and, beyond CFD, turn LES into a valuable tool for fundamental turbulence research.
\end{abstract}

\section{Introduction}
\label{sec:1}

Large-eddy simulations are used in applied studies to simulate large-scale turbulent dynamics, relying on empirical modeling at small (subgrid) scales \citep{pope2000,meneveau2000scale,sagaut2005large,fureby2008}. They are capable of generating intermittent (multifractal) velocity fields $\bu(\bx,t)$, reproducing key features of homogeneous 3D turbulence \citep{cerutti1998intermittency,biferale2019self,linkmann2018multi}.
Classical description of intermittency involves, in particular, the distributions of longitudinal velocity increments $\Delta_\ell u_{\parallel} = \big[ \bu(\bx +\ell \bee) - \bu(\bx) \big] \cdot \bee$ \corr (with $\bee$ arbitrary unit vector) \rroc becoming increasingly flat-tailed at small scales and leading to anomalous scaling of structure functions \citep{frisch1995turbulence,chevillard2015peinture,iyer2017reynolds};
 see  Fig.\ref{fig:1}.
 The multifractal phenomenology is rooted in the scaling symmetries of the Navier-Stokes (NS) equations, suggesting that these symmetries should drive any modeling strategies of turbulence including LES. 
\begin{figure}
\begin{center}
\includegraphics[width=0.335\textwidth]{1a.png}
\includegraphics[width=0.5\textwidth]{1b.pdf}
\end{center}
\caption{(a) 3D rendering of the velocity norm $|\bu|$ in units of the \corr root-mean-squared velocity $u_{rms} \simeq 2.4$  \rroc for LES at resolution $N^3 = 1024^3$. (b) Broken self-similarity. Shown is the flatness of the longitudinal velocity increments $\Delta_\ell u_{\parallel}$ for different resolutions \corr in  units of $L=\pi$ and  with forcing scale $L_f =4L/3$\rroc; see the \emph{implicit} line in Tab. \ref{table:1} for numerical details. For reference, in purple we show the anomalous scaling $\propto \ell^{-0.12}$.
The inset shows \corr their  PDFs normalised by their standard deviation $\sigma(\Delta_\ell u_\parallel)$ \rroc (shifted vertically for better visibility) for the $1024^3$ run at scales ranging geometrically from $\delta$  to $128 \delta$ from top to bottom. 
}
\label{fig:1}
\setlength{\unitlength}{\columnwidth}
\begin{picture}(1,0)(0,0)
\put(0.235,0.63){(a)}
\put(0.67,0.63){(b)}
\end{picture}
\end{figure}
Here we investigate the Hidden Scale Invariance (HS) approach developed recently by \cite{mailybaev2020hiddena,MailybaevTransactions2022}, which describes scaling properties of the NS in a quasi-Lagrangian reference frame \citep{belinicher1987scale,Mitra,biferale2011multi} after dynamic (nonlinear) rescaling. By revealing unbroken (hidden) self-similarity of  intermittent statistics, this approach provides theoretical support to the anomalous scaling of structure functions \citep{mailybaev2022shell,thalabard2024zero,calascibetta2025HS} and the universal statistics of multipliers, the ratios like $\Delta_{\ell} u_{\parallel} / \Delta_{\ell'} u_{\parallel}$, 
\corr as conjectured by Kolmogorov in his refined theory of turbulence \rroc  \citep{kolmogorov1962refinement,chen2003kolmogorov}. 
Extension of the HS to SGS closures has been recently studied in shell models \citep{mailybaev2020hiddenb}. 

In this work, we prove that properly designed SGS models, for which the Smagorinsky is the simplest example, preserve the HS. This implies that the rescaled quasi-Lagrangian statistics does not depend on the observation scale, $\ell$,  and the filtering (cut-off) scale,  $\delta$, separately but only through their ratio $\alpha = \ell/\delta$. 
\corr Besides, for those resolved scales $ \delta \ll \ell \ll L_f$ 
far from the cut-off and the forcing scale $L_f$, e.g.  in the inertial range,
the statistics becomes independent of $\alpha$ and coincides with the analogous hidden-symmetric statistics of the NS. \rroc Performing high-resolution simulations, we provide very accurate numerical verification of HS. Such accuracy is attained because the HS in SGS models extends to subgrid scales, unlike the NS where it only refers to the inertial range. 
This stronger form of HS has implications for both practical (development of SGS models) and theoretical studies of turbulence.

\section{Kinematic rescaling of a quasi-Lagrangian velocity}
\label{sec:2}
\subsubsection*{Navier-Stokes-Smagorinsky equation}
In practice, the closure problem takes as inputs a filtering scale $\delta$ and  a filter $G$. It aims to predict the dynamics of the filtered field as the convolution
$\fil{\bu}{\delta} = G_\delta * \bu$ with the kernel $G_\delta(\bx) = \delta^{-3} G(\bx/\delta)$, providing implicit or explicit parametrizations  for  the so-called subgrid scale tensor $\bR_\delta =\fil{\left(\bu \otimes \bu\right)}{\delta}-\fil{\bu}{\delta}\otimes\fil{\bu}{\delta}$.
We start with the simplest example of the Smagorinsky model, in which $\bR_\delta$ is expressed as the viscous stress with the effective eddy-viscosity $\nu_s$. Dropping $\delta$'s from the subscripts, the resulting Navier-Stokes-Smagorinsky (NSS) dynamics for the filtered velocity becomes
\be
   \label{eq:NSS}
    \partial_t  \bu  +\bu \cdot \bnabla \bu = -\bnabla p+\bnabla \cdot \left(2 \nu_s \bs\right)+\bff, \quad 
  	\bnabla \cdot \bu = 0,
\ee
where $p(\bx,t)$ is the pressure and $\bff(\bx,t)$ is a large-scale force with correlation length $L_f\gg \delta$. 
The eddy viscosity is defined in terms of the filtered rate-of-strain tensor $\bs$ as
    \be
    \label{eq:nu_s}
    \nu_s = (c_s \delta)^2 \|\bs\|,\quad  \bs 
    = \frac{1}{2} \left( \bnabla \bu + (\bnabla \bu)^T\right),
    \ee
with the Smagorinsky constant $c_s = 0.1$ and the norm $\|\bs\|^2 = \sum 2 s_{ij}^2$. 
Considering the filtering scale $\delta$ to be much larger than the viscous scale, the kinematic viscosity $\nu \ll \nu_s$ is small and, therefore, omitted in the NSS equations (\ref{eq:NSS}). 
The filtering operation leading to Eq.~\eqref{eq:nu_s} is formal only. While the  NSS dynamics aims to mimic the Navier-Stokes (NS) equations for the resolved scales $\ell \gg \delta$, the role of eddy viscosity is to replace the energy transfer at the subgrid scales $\ell \lesssim \delta$ by a dissipative scheme \citep{cruz2023physical}. 

\subsubsection*{Kinematic rescaling}
 Similarly to the NS system previously discussed by \cite{mailybaev2020hiddena,MailybaevTransactions2022}, 
the starting point of our study is the projection of the velocity field $\bu(\bx,t)$ into the rescaled velocity field $\bU(\bX,\tau)$.
At a heuristic level, the new field represents a quasi-Lagrangian velocity increment \citep{belinicher1987scale} with a dynamically rescaled amplitude and timeframe. The course of proper time $\tau$ is determined by the local level of fluctuations,  running fast when the latter are high and slowly otherwise. 
The rescaling  is parametrized by an observational  scale $\ell$ and a label $\bx_0 \in \mathbb{R}^3$. The label selects a Lagrangian trajectory as
\begin{equation}
	\label{eq:kinLT}
	 \dfrac{d\bx_*}{dt} = {\bu}(\bx_*,t),\quad \bx_*(0)= \bx_0.
\end{equation}
Fluctuations at scale $\ell$ are measured as amplitudes of quasi-Lagrangian increments as  
	\begin{equation}
	\label{eq:kin2}
	 a_\ell(t) = \sqrt{\int_{|\mathbf{X}| \le 1} 
     \big|\Delta_\ell \bu\big(\bx_*(t),\bX,t\big) \big|^2 d^3\bX}, \quad 
	\Delta_\ell \bu(\bx,\bX, t) = \bu\big(\bx+\ell \bX,t\big)-\bu\big(\bx,t\big).
	\end{equation}
%
\corr The space-variable $\bX$ represents the  vectorial  separation with respect to the Lagrangian trajectory in units of the observational scale $\ell$.
\rroc  
The velocity field is then rescaled  as
\be
	\label{eq:kin3}
	 \bU(\bX,\tau) = \dfrac{\Delta_\ell\bu\big(\bx_*(t),\bX,t\big)}{a_\ell(t)} ,\quad
	 \tau = \frac{1}{\ell}\int_0^t a_\ell(s) \, ds.
\ee
\corr
Here the first formula introduces the rescaled velocity $\bU$ depending on the rescaled time $\tau$, which is expressed in terms of the velocity difference $\Delta_\ell\bu$ at original time $t$. The second formula defines the relation between $t$ and $\tau$, which is one-to-one for $a_\ell(t) > 0$. 
\rroc
Physically, $\tau$ represents a Lagrangian clock measuring time in units of the local turnover time $ \ell/a_\ell(t)$ encountered along the trajectory $\bx_*(t)$.


\corr
Note that the choice of $a_\ell$ in Eq.(\ref{eq:kin2}) is by no mean unique. In practice, the essential requirement is that it should be homogeneous of degree 1 with respect to the velocity field and positive along a trajectory. Here, for simplicity, we restrict our discussion to a specific example rather than deriving the hidden symmetry in full generality. We refer to \cite{MailybaevTransactions2022} for more general derivations.
\rroc

\subsubsection*{Switching the observation scale}
Changing the scale $\ell \mapsto \ell'$ in Eqs.~(\ref{eq:kin2}) and (\ref{eq:kin3}) reduces to a specific transformation of the respective fields $\bU(\bX,\tau) \mapsto \bU'(\bX,\tau')$ expressed as 
\be
	\label{eq:ell}
	\bU'(\bX,\tau')  = \frac{\bU(\bX/\lambda,\tau)}{A(\tau)}, \ \
	\tau' = \lambda \int_0^\tau  A(s) \, ds, \ \
	A(\tau) = \sqrt{\int_{|\mathbf{X}| \le 1} | \bU(\bX/\lambda,\tau) |^2 d^3\bX},
\ee
where $\lambda = \ell/\ell'$ and the second relation introduces the change of time $\tau \mapsto \tau'$. 
The kinematic relations (\ref{eq:kin3}) and (\ref{eq:ell}) are demonstrated schematically in Fig.~\ref{fig:2}(a) by straight and wavy arrows, respectively. In this figure, we specified the parameters in the subscripts: $\bu_\delta$ for the original field depending on the filtering scale $\delta$ and $\bU_{\delta,\ell,\bx_0}$ for the rescaled field depending additionally on the observation scale $\ell$ and the label $\bx_0$.

\begin{figure}
\centering
\includegraphics[width=0.7\textwidth]{2}
\caption{(a) Kinematic rescaling diagram, where the projections (\ref{eq:kin3}) are shown by straight arrows for two different scales $\ell$ and $\ell' = \lambda\ell$, and the transformation (\ref{eq:ell}) is shown by the wavy arrow. (b) Similar diagram for a different filtering scale $\delta' = \delta/\lambda$. (c) Manifestation of the hidden self-similarity as the fusion of the previous diagrams in their right-hand sides.}
\label{fig:2}
\end{figure}

\subsubsection*{NSS under kinematic rescaling}
The governing equations for the rescaled velocity field $\bU(\bX,\tau)$ are obtained by differentiating the definition (\ref{eq:kin3}) with respect to $\tau$ and using the NSS system (\ref{eq:NSS}) and (\ref{eq:nu_s}). The lengthy but elementary derivation yields (see \S\ref{sec:appendixA})
	\begin{equation}
	\label{eq:HSmago}
	\partial_\tau \bU 
	= \Lambda_\bU \big[-\bU \cdot \bnabla \bU-\bnabla P+2 c_s^2 \,\alpha^{-2}\, \bnabla \cdot (\|\bS\| \bS)+\bfF \big], \quad
	\bnabla \cdot {\bU}=0.
	\end{equation}
Here $\alpha = \ell/\delta$, the tensor $\bS = \frac{1}{2} \bnabla \bU + \frac{1}{2}(\bnabla \bU)^T$, and the gradient operator $\bnabla$ acts on the new space variable $\bX$. The operator $\Lambda_\bU$ is defined for any field $\bV(\bX,\tau)$ as
	\begin{equation}
	\label{eq:HSL}
	\Lambda_\bU[\bV] = \tilde{\bV}-\bU \int_{|\mathbf{X}| \le 1} 
	\bU \cdot \tilde{\bV} d^3\bX, \quad
	\tilde{\bV} = \bV(\bX,\tau)-\bV(\mathbf{0},\tau).	 
	\end{equation}
The rescaled forcing term is expressed, with the times $t \mapsto \tau$ related by Eq.~(\ref{eq:kin3}), as
	\begin{equation}
	\label{eq:Fnu}
	\bfF(\bX,\tau) 
    =  \frac{\ell \, 
    \Delta_\ell \bff\big(\bx_*(t),\bX,t\big)}{a_\ell^2(t)}, \quad
	\Delta_\ell \bff(\bx,\bX,t) = \bff(\bx+\ell \bX,t)-\bff(\bx,t).
	\end{equation}

\section{Hidden scale invariance}
\subsubsection*{The limit of small observational scales}
System~(\ref{eq:HSmago}) governs the dynamics of the new velocity and pressure fields $\bU(\bX,\tau)$ and $P(\bX,\tau)$.
It is, however, not closed with respect to these fields, because the coupling with the original dynamics $\bu(\bx,t)$ permeates through the dependence of  forcing term (\ref{eq:Fnu}) on $a_\ell(t)$. %
We now argue that a closed formulation follows asymptotically at small observational scales $\ell \ll L_f$, in which case the forcing term is negligible. Indeed, we have $\partial_\tau \bU \sim 1$ for $\bX \sim 1$ by construction from Eqs.~(\ref{eq:kin2}) and (\ref{eq:kin3}). 
This implies that the  lhs in Eq.~\eqref{eq:HSmago} is $\sim 1$. As for the forcing term, we first consider the case $\delta \lesssim \ell \ll L_f$,
describing an observational scale in the resolved (turbulent) range.
The dimensional estimate $\bff \sim U_f^2/L_f$ in Eq.~(\ref{eq:Fnu}) yields $\Delta_\ell\bff \sim U_f^2 \ell/L_f^2$, with 
$U_f$  the integral (forcing-scale) velocity. Using the K41 estimate $a_\ell \sim U_f (\ell/L_f)^{1/3}$ for the  amplitude of velocity fluctuations in Eq.~(\ref{eq:Fnu}), we conclude that the forcing term $\bfF  \sim (\ell/L_f)^{4/3} \ll 1$. This estimate extends to the subgrid range  $\ell <\delta \ll L_f$ as $\bfF  \sim (\delta/L_f)^{4/3} \ll 1$. 

Therefore, under the condition of small scales $\delta, \ell \ll L_f$, the forcing term can be neglected in the rescaled system \eqref{eq:HSmago}. This gives
	\begin{equation}
	\label{eq:HSmagoA}
	\delta, \ell \ll L_f: \quad
	\partial_\tau \bU 
	= \Lambda_\bU\big[-\bU \cdot \bnabla \bU-\bnabla P+ 2 c_s^2 \alpha^{-2} \bnabla \cdot (\|\bS\| \bS) \big], \quad
	\bnabla \cdot {\bU}=0.
	\end{equation}
We call the system \eqref{eq:HSmagoA} $\alpha$-NSS, emphasizing that the dependence on the parameters $\ell$ and $\delta$ is realized only through their ratio $\alpha = \ell/\delta$. Based on these properties, we formulate the hypothesis of \textit{hidden scale invariance}:
\vspace{2mm}
\begin{center}
For small scales $\ell,\delta \ll L_f$, the statistics of the rescaled field $\bU(\bX,\tau)$, obtained \\
from the NSS solution $\bu(\bx,t)$, depend on $\ell$ and $\delta$ only through their ratio $\alpha = \ell/\delta$. 
\end{center}

\subsubsection*{Statistical identities}
The diagrammatic interpretation of this type of  scale invariance is presented in Fig.~\ref{fig:2}. The panels (a) and (b) show two independent diagrams corresponding to different filtering parameters $\delta$ and $\delta' = \lambda\delta$. When viewed at small scales and in a statistical sense, these diagrams merge on their right sides as shown in the panel (c), where $\bU_\alpha$ denotes a solution of the $\alpha$-NSS system~(\ref{eq:HSmagoA}). In other terms, the hidden scale invariance is understood as the statistical equivalence 
	\begin{equation}
	  \bU_{\delta,\ell,\bx_0} 
	\stackrel{\mathrm{law}}{=} \, \bU_{ \lambda \delta,\lambda \ell ,\bx'_0} \stackrel{\mathrm{law}}{=}  \bU_\alpha \ \
	\textrm{for} \ \  \alpha = \ell/\delta.
 	\label{eq:HSU}
	\end{equation}
The diagram in Fig.~\ref{fig:2}(c) also reveals that the rescaled statistics for different values of $\alpha$ and $\alpha' = \alpha/\lambda$ are related by the transformation $h_\lambda: \bU_\alpha \mapsto \bU_{\alpha'}$ prescribed by Eq.~(\ref{eq:ell}). 
This provides an additional statistical identity
	\begin{equation}
	\label{eq:HSU_2}
	  \bU_{\alpha/\lambda}  \stackrel{\mathrm{law}}{=} h_\lambda [\bU_\alpha].
	\end{equation}

\subsubsection*{HS in the inertial interval}
A stronger form of HS develops in the inertial interval, i.e., at scales $\ell$ defined by the condition $\delta \ll \ell \ll L_f$. 
Here, the Smagorinsky dissipative term becomes negligible and the system (\ref{eq:HSmagoA})  simplifies into
	\begin{equation}
	\label{eq:HSmagoB}
	\delta \ll \ell \ll L: \quad
	\partial_\tau \bU + \Lambda_\bU \big[\bU \cdot \bnabla \bU+\bnabla P \big]=0, \quad 
    \bnabla \cdot {\bU}=0.
	\end{equation}
This system is independent of both $\delta$ and $\ell$, and coincides with the rescaled system derived earlier in the framework of NS turbulence by \cite{mailybaev2020hiddena,MailybaevTransactions2022}. 
This leads to the hypothesis of \textit{inertial-interval hidden symmetry}: 

\vspace{2mm}
\begin{center}
	The statistics of the rescaled field $\bU(\bX,\tau)$ in the inertial interval are universal, \\ 
	i.e. independent of the scale $\ell$ and the dissipation mechanism.
\end{center}

\vspace{2mm} \noindent 
This suggests that, despite the intermittency, the rescaled NSS and NS statistics coincide at the inertial interval scales.
The statistical identity \eqref{eq:HSU_2} in the inertial interval simplifies into
	\begin{equation}
	\label{eq:HSU_3}
	\bU_\infty  \stackrel{\mathrm{law}}{=} h_\lambda [\bU_\infty],
	\end{equation}
where $\bU_\infty$ denotes a solution of system (\ref{eq:HSmagoB}) obtained in the limit $\alpha \to \infty$. 

\subsubsection*{HS in other SGS models}
We observe that the concept of hidden scale invariance applies equally to hyperviscous Smagorinsky models, such as $\propto \! \delta^{2p} |\bnabla^2|^p \bnabla \cdot(\nu_s \bs)$.
More interestingly, modern strategies in LES simulations use implicit dissipation schemes in order to introduce backscattering effects; see, \emph{e.g.}, \cite{pope2000,lamballais2021viscous}.
In practice, these schemes are implemented by under-resolving the NSS  through a (genuine) filtering operation, hereby
transforming the NSS equations    \eqref{eq:NSS} into
\be
   \label{eq:NSS-filtered}
    \partial_t  \bu 
    = \big(-\bu \cdot \bnabla \bu-\bnabla p+\bnabla \cdot \left(2 \nu_s \bs\right)+\bff\big)_{\delta}, \quad 
  	\bnabla \cdot \bu = 0,
\ee
where the rhs is the filtered field. Consider the sharp spectral filter $G_\delta(\bx) = (2\pi)^{-3}\int_{|\bk|\le c k_\delta} \! d^3 \bk e^{i\bk\cdot \bx}$ defined as a Galerkin truncation at a given multiple $c$ of the wavenumber $k_\delta = 2\pi/\delta$. Then the kinematic rescaling  of \S \ref{sec:2} yields the filtered $\alpha$-NSS as 
	\begin{equation}
	\label{eq:HSmago-filtered}
	\partial_\tau \bU 
	= \big(\Lambda_\bU [
	-\bU \cdot \bnabla \bU-\bnabla P+2c_s^2 \alpha^{-2}\, \bnabla \cdot (\|\bS\| \bS)+\bfF ]\big)_{\alpha},
	\quad \bnabla \cdot {\bU} =0
	\end{equation}
(see \S\ref{sec:appendixB}). Here the filtering operation $(\dots)_\alpha$ applies wrt the rescaled variable $\bX$ at scale $\alpha = \ell/\delta$.
Hence, the HS extends to implicit LES schemes. Following \cite{mailybaev2020hiddena} one can relate the HS of SGS models to their invariance with respect to space-time scaling and Galilean symmetry groups. From this point of view, the extension of HS to subgrid (dissipative) scales refers to the time-scale-invariant (quadratic in velocity) SGS model. On the contrary, the HS is limited to the inertial scales of the NS, because physical viscosity breaks the time-scale invariance \citep{MailybaevTransactions2022}.

\section{Numerical verification of the hidden scale invariance}
\subsubsection*{Numerical setup}
We probe the hidden self-similarity conjectures through direct numerical simulations of the NSS
in a triply $2\pi$-periodic domain.  Our simulations use  a pseudospectral scheme with 2/3-rule dealiasing  \citep{orszag1971elimination} and a second-order Adam-Bashforth  time integration with step $\Delta t$. 
The forcing $\bff(\bx,t) =  \bnabla \times \psi$ is specified through the stream vector
$\bold \psi= \sqrt{\epsilon} \sum_{ 1 \le |\bk| \le 2\pi/L_f} e^{i \bk \cdot \bx} {\boldsymbol\eta (\bk,t)}/(\sum_{1 \le |\bk| \le 2\pi/L_f} |\bk|^2)$.
Here $\epsilon = 0.14$ and $L_f = 4\pi/3$ are the energy injection rate and scale, and $\boldsymbol \eta(\bk,t)$ are white-in-time independent complex standard Gaussian random vectors satisfying the Hermitian symmetry $\boldsymbol\eta(\bk)=\boldsymbol\eta^*(-\bk)$. 
We describe two sets of runs with resolution $N$ ranging from $256^3$ to $1024^3$,  referred to as  explicit and implicit runs.
In the explicit runs use $\delta \gg \Delta x$ ensuring that dissipation in the subgrid range is dominated by the Smagorinsky damping term, while the implicit runs use $\delta \sim \Delta x$ and allow for energy backscatter; see Table \ref{table:1} for more details. For numerical convenience we use $|\bX| = \max_j|X_j|$ in the amplitudes (\ref{eq:kin2}) corresponding to averages in a cube.

\begin{table}
\centering
\begin{tabular}{ccccc}
& $\delta$ & $ck_\delta$ & $c$ & $\Delta t$    \\
\hline
explicit & $9.6\, \Delta x$ & $k_{\max}$ & 3.2 & $0.15/N$  \\
\hline
implicit & $1.8\, \Delta x$ & $k_{\max}$  & 0.6 & $0.10/N $  
\end{tabular}
\caption{Numerical details for two sets of runs with $k_{\max}=N/3$ and $\Delta x = 2\pi/N$ using the filtered NSS Eq.~\eqref{eq:NSS-filtered} at resolutions $ N^3 = 256^3, 512^3, 1024^3$. 
 The total integration time is $T_{\max} = 15$ for resolutions up to $512^3$ and $T_{\max}=4.6$ for $1024^3$. }
\label{table:1}
\end{table}

\subsubsection*{Temporal averages}
The statistics of the rescaled field $\bU(\bX,\tau)$ can be accessed using different observables. In this work we limit our analysis to the longitudinal component $U_\parallel(\tau) = \bU(\bee,\tau)\cdot \bee$ with the unit vector $\bee = (1,0,0)$; other observables (not presented here) were also tested with similar conclusions.
The $\tau$-average of $|U_\parallel|^p$ is defined as
\begin{equation}
	\label{eq:phi}
	\av{|U_\parallel(\tau)|^p}_\tau=\lim_{T\to \infty}\dfrac{1}{T}\int_0^T |U_\parallel(\tau)|^p d\tau,
\end{equation}
from which the usual statistical properties such as variance, skewness, flatness, etc. are estimated.
The corresponding probability density function (PDF) is obtained from the average of the Dirac delta function as $p_{\parallel}(y)= \av{\delta\left(U_\parallel(\tau)-y\right)}_\tau$.
We emphasize that the average \eqref{eq:phi} is expressed in terms of the rescaled Lagrangian clock $\tau$. However, 
it can be efficiently estimated as space-time average over the original fields  using a conversion formula relying on  spatial homogeneity and  incompressibility. For $|U_\parallel|^p$ the conversion formula states
\be
	\label{eq:PDF}
	\av{|U_\parallel(\tau)|^p}_\tau = \av{\left |\dfrac{\Delta_\ell u_\parallel(\bx,t)}{a(\bx,t)}\right |^p\, J(\bx,t)}_{\bx,t},
\ee
where $\av{\cdot}_{\bx,t}$ denotes the space-time average in the original  coordinates and $\Delta_\ell u_\parallel = \big( \bu(\bx +\ell \bee) - \bu(\bx) \big) \cdot \bee$. 
\corr
In other words, Eq.~\eqref{eq:PDF} transform quasi-Lagrangian averages into usual (Eulerian) ones, that one can easily compute without having to track any fluid trajectory.
\rroc
Here $J(\bx,t)$ is the Jacobian factor due to the change of time $\tau \mapsto t$ as
\begin{equation}
	J(\bx,t) = \frac{a(\bx,t)}{\av{a(\bx,t)}_{\bx,t}}, \quad
	a(\bx,t) = \Big(\int_{|\mathbf{X}| \le 1} \big| \Delta_\ell\bu(\bx,\bX,t) \big|^2 d^3\bX\Big)^{1/2}.
\end{equation}
We refer to \cite{MailybaevTransactions2022} for technical details of this derivation. 
Expressions similar to Eq.~(\ref{eq:PDF}) follow for other single-time observables.

\subsubsection*{Numerical results}
\corr
Figure~\ref{fig:3}(a) illustrates the HS by verifying the collapse at the level of PDFs of $U_\parallel$ for increasing prescribed values   of the parameter $\alpha = \ell/\delta$. The top and bottom panels correspond to the explicit and implicit runs, respectively, and we recall that different resolutions feature different $\delta$; see Tab.~\ref{table:1}. In both  cases, the curves collapse at small    scales, verifying the hidden self-similarity: the statistics depend only on the ratio $\ell/\delta$. 
As $\alpha$ increases,  the graphs begin to diverge from one another, with the smallest-resolution run departing from the other two.
This signals a  sensitivity of the statistics to the large (forcing) scales, and this is compatible with the HS getting broken at scales $ O(L_f/\delta)$.

Figure~\ref{fig:3}(b) further characterises this large-scale sensitivity, by showing the  variance and flatness  of $U_\parallel(\tau)$ as functions of $\alpha = \ell/\delta$. Looking at the statistics from large to small scales (right to left), the gradual convergence is most visible at the level of the flatness, signaling that the PDF tails require a larger inertial range to converge ---see also the insets in Fig.~\ref{fig:3}(a). 
\rroc
\begin{figure}
    \centering
    \includegraphics[width = 0.65\textwidth,trim=0cm 1.7cm 0cm 0cm,clip]{3a_exp.pdf}
    \includegraphics[width = 0.34\textwidth,trim=0cm 1.7cm 0cm 0cm,clip]{3b_exp.pdf}\\
    \vspace{0.5cm}
    \includegraphics[width = 0.65\textwidth]{3a_imp.pdf}
    \includegraphics[width = 0.34\textwidth]{3b_imp.pdf}
%
    \caption{ \corr (a) PDF of $ U_\parallel$ at selected values of $\ell/\delta$ for the various runs. The data is shifted horizontally for clarity with the  vertical lines indicating the shifts. The top insets show the corresponding data using logarithmic vertical axis. (b) Variance and flatness of $U_\parallel$ as a function of $\alpha = \ell/\delta$.  The vertical lines indicate the values of $\alpha$ used in the left panel. Top and bottom panels use, respectively, the explicit and implicit simulations. \rroc}
    \label{fig:3}
\setlength{\unitlength}{\columnwidth}
\begin{picture}(1,0)(0,0)
\put(0.07,0.81){(a)}
\put(0.7,0.81){(b)}
\put(0.2,0.49){$N:$ \color{ForestGreen} $--$ \color{black} $256^3$  \color{blue} $--$ \color{black} $512^3$ \color{black} $-$ \color{black} $1024^3$}
\end{picture}
\end{figure}
Explicit runs use large $\delta$ and therefore do not develop the inertial range; their HS applies to dissipative scales only.
Implicit runs have smaller $\delta$, and a pronounced inertial interval develops at the resolution $1024^3$ for the scales $6 \lesssim \ell/\delta \lesssim 42$; \corr see the plateau in the bottom Fig.~\ref{fig:3}(b) for both the variance and the flatness.
\corr
 Note that the flatness curves do not collapse well in this range because of  the shorter inertial interval for lower resolutions. 
\rroc
PDFs of $U_\parallel$ for 18 different scales $\ell$ from the inertial interval are shown by solid lines in Fig.~\ref{fig:4}(a); \corr  these scales $0.01<\ell/L<0.1$ correspond to the squares in Fig.~\ref{fig:4}(b). They also correspond to the range over which a flatness scaling $\propto \ell^{-0.12}$ is clearly observable in Fig.~\ref{fig:1}(b).
In Fig.~\ref{fig:4}(a), these  PDFs are visually indistinguishable, confirming the inertial-interval HS. \rroc
The collapse is quantified in Fig.\ref{fig:4}(b), where the black  squares show the $L_2$-distance between the PDF at scale $\ell$ and reference PDF from the middle of the inertial interval.

\begin{figure}
    \centering
    \includegraphics[width = 0.395\textwidth]{4a.pdf}
    \hspace{5mm}
    \includegraphics[width = 0.395\textwidth]{4b.pdf}
    \caption{(a) Solid lines represent 18 PDFs of $U_\parallel$ for the $1024^3$ implicit run; see full black square in the right panel.
    Red dashed lines present 6 PDFs of $U_\parallel$ for the JHU $8192^3$ NS simulation; see full red circles in the right panel. 
 (b) $L_2$-distance between the PDFs of the left panel and the reference NSS PDF from the middle of the inertial interval. }
    \label{fig:4}
\setlength{\unitlength}{\columnwidth}
\begin{picture}(1,0)(0,0)
\put(0.133,0.42){(a)}
\put(0.585,0.415){(b)}
\end{picture}
\end{figure}

We further test the universality of inertial-interval HS with respect to the dissipation mechanism. For this purpose, we superpose our results with the data obtained from the NS simulations of the JHU turbulence database at $N = 8192^3$\citep{li2008public,yeung2012dissipation,yeung2015extreme}. The 6 undistinguishable dashed red lines in Fig.\ref{fig:4}(a) are the PDFs of $U_\parallel$ for different scales from the inertial interval; see the full red dots in the right panel. The close-to-perfect superposition of the PDFs for the NSS and NS simulations confirms the universality of inertial-interval HS. Red circles in Fig.~\ref{fig:4}(b) quantify this collapse by measuring $L_2$-distance between the NS PDF at scale $\ell$ and the same reference PDF from the NSS data. It is remarkable that the NS inertial interval is shorter despite of much higher resolution.


\section{Concluding remarks}
\label{sec:5}

Similar to the NS case,  the hidden scale invariance of classical LES  models  involves a quasi-Lagrangian formulation, in which the timeframe is adjusted dynamically to an amplitude of local fluctuations at a scale $\ell$.
Statistics of the resulting field  depend on the observation scale $\ell$ and on the filtering scale $\delta$ only through their ratio $\ell/\delta$. They become universal in the inertial interval, i.e. independent of both $\ell$ and $\delta$ and coincident with the analogous statistics of 3D Navier-Stokes turbulence.
The validity of the hidden scale invariance at both inertial and dissipative scales in the SGS models, in contrast to its restriction to inertial scales only for the NS, allows a very accurate numerical verification of the hidden self-similarity and makes LES useful for the theoretical study of intermittency in turbulence. 
\corr In particular, we have in mind the potential use of Smagorinsky (or variants) as a model where one can explain the anomalous scaling in terms of the hidden symmetry, as was done earlier in shell models~\citep{mailybaev2022shell,mailybaev2020hiddenb,thalabard2024zero} and 2D passive scalar turbulence \citep{calascibetta2025HS}. 
\rroc
The hidden self-similarity across the entire small-scale range may also have applications to the design and optimization of SGS models; see \citep{Biferale2017,domingues2024data,freitas2025posteriori} for similar studies in shell models. 



\backsection[Funding]{This work was supported by the European Research Council (ERC) under the European Union’s Horizon 2020 research and innovation program (Grant 882340), CNPq grants 308721/2021-7 and 150944/2024-1, FAPERJ grant E-26/201.054/2022,  CAPES grant AMSUD3169225P, and the Brazilian-French Network in Mathematics.}

\backsection[Declaration of interests]{The authors report no conflict of interest.}

\backsection[Data availability statement]{The data that support the findings of this study are openly available in HS LES Turbulence at http://doi.org/10.5281/zenodo.15376629, reference number 15376629. See JFM's \href{https://www.cambridge.org/core/journals/journal-of-fluid-mechanics/information/journal-policies/research-transparency}{research transparency policy} for more information}

\appendix
\section{Appendix}
\subsection{Derivation of the rescaled system (\ref{eq:HSmago})}
\label{sec:appendixA}
For technical derivations, we drop the subscript $\ell$ and use the shorthand for Eq.~(\ref{eq:kin2}) as
	\begin{equation}
	 a(t) = \sqrt{\int_{|\mathbf{X}| \le 1} \bv \cdot \bv \, d^3\bX}, 
     \quad
	\bv(\bX, t) 
    = \bu\big(\bx_*(t)+\ell \bX,t\big)-\bu\big(\bx_*(t),t\big).
	\label{eqA0}
	\end{equation}
Differentiating the first equality in Eq.~(\ref{eq:kin3}) with $d\tau = a(t)dt/\ell$ yields
	\begin{equation}
	\partial_\tau \bU
	= \frac{\ell}{a} \frac{\partial}{\partial t} \frac{\bv}{a}
	= \frac{\ell}{a^2} \frac{\partial \bv}{\partial t}-\frac{\ell \bv}{a^3} \frac{d a}{d t} 
	= \frac{\ell}{a^2} \frac{\partial \bv}{\partial t}
	-\frac{\bv}{a} \int_{|\mathbf{X}| \le 1} \frac{\bv}{a} \cdot 
	\left(\frac{\ell}{a^2}\frac{\partial \bv}{\partial t}\right) d^3\bX,
	\label{eqA1}
	\end{equation}
where we expressed $da/dt$ from Eq.~(\ref{eqA0}) and rearranged the factors. Differentiating the second expression in Eq.~(\ref{eqA0}) with the chain rule and Eq.~(\ref{eq:kinLT}) yields
	\begin{equation}
	\begin{array}{l}
	\displaystyle
	\frac{\partial \bv}{\partial t} 
	= 
	\bu\big|_{\bx_*(t),t} \cdot \big(\bnabla\bu \big) \big|^{\bx_*(t)+\ell \bX,t}_{\bx_*(t),t} 
	+\frac{\partial \bu}{\partial t}  \Big|^{\bx_*(t)+\ell \bX,t}_{\bx_*(t),t}
	= 
	\bu\big|_{\bx_*(t),t} \cdot \big(\bnabla\bu \big) \big|^{\bx_*(t)+\ell \bX,t}_{\bx_*(t),t} 
	\\[12pt]
	 \displaystyle \qquad \ \ +\,
	\big(- \bu \cdot \bnabla\bu  -\bnabla p+2(c_s \delta)^2 \bnabla \cdot \left(\|\bs\| \bs\right) +\bff \big) \big|^{\bx_*(t)+\ell \bX,t}_{\bx_*(t),t},
	\end{array}
	\label{eqA2}
	\end{equation}
where the last equality used the time derivative from Eqs.~(\ref{eq:NSS}) and (\ref{eq:nu_s}).
Using Eqs.~(\ref{eqA0}) and (\ref{eq:kin3}) one can check that 
	\begin{equation}
	\bu\big|_{\bx_*(t),t} \cdot (\bnabla \bu)\big|^{\bx_*(t)+\ell\bX,t}_{\bx_*(t),t}
	-(\bu \cdot \bnabla\bu) \big|^{\bx_*(t)+\ell\bX,t}_{\bx_*(t),t}
	= -\frac{1}{\ell} \, (\bv \cdot \bnabla_\bX \bv)\big|^{\bX,t}_{\mathbf{0},t},
	\label{eqA2f}
	\end{equation}
where we distinguished the gradient $\bnabla_\bX$ in the $\bX$ space. Similarly, 
	\begin{equation}
	\left[ \bnabla \cdot \big( \|\bs\| \bs \big)\right]^{\bx_*(t)+\ell \bX,t}_{\bx_*(t),t}
	= \frac{1}{\ell^3} \left[  \bnabla_{\bX} \cdot \big( \|\tilde{\bs}\| \tilde{\bs} \big)\right]_{\mathbf{0},t}^{\bX,t},  
	\label{eqA4b}
	\end{equation}
for the tensors $\bs = \frac{1}{2} \bnabla \bu + \frac{1}{2}(\bnabla \bu)^T$ and $\tilde{\bs} = \frac{1}{2} \bnabla_\bX \bv + \frac{1}{2}(\bnabla_\bX \bv)^T$.
One can check now that the rescaled system (\ref{eq:HSmago}) with the operator (\ref{eq:HSL}) follows from Eq.~(\ref{eqA1}) after the substitutions of $\bv/a = \bU$ and the time derivative (\ref{eqA2}) with the terms expressed from Eqs.~(\ref{eqA2f}), (\ref{eqA4b}), the rescaled pressure $P(\bX,\tau) = p\big(\bx_*(t)+\ell \bX,t\big)/a^2(t)$ and the forcing term (\ref{eq:Fnu}).

\subsection{Derivation of the rescaled system (\ref{eq:HSmago-filtered})}
\label{sec:appendixB}

Finally, let us consider a velocity  satisfying  the  NSS \eqref{eq:NSS-filtered} convolved by the sharp spectral filter  $G_\delta(\bx) = (2\pi)^{-3}\int_{|\bk|\le ck_\delta} d^3 \! \bk e^{i\bk\cdot \bx}$ with $k_\delta = 2\pi/\delta$.
The steps yielding the rescaled dynamics in Eq.~(\ref{eq:HSmago-filtered}) are similar, but now one should additionally deal with the filtered terms. 
One can see that Eqs.~(\ref{eqA0}) and (\ref{eqA1}) remain valid, while the last term in Eq.~(\ref{eqA2}) gains the filter $(...)_\delta\, \big|^{\bx_*(t)+\ell \bX,t}_{\bx_*(t),t}$. Next one can check that Eq.~(\ref{eqA2f}) becomes
	\begin{equation}
	\bu\big|_{\bx_*(t),t} \cdot (\bnabla \bu)\big|^{\bx_*(t)+\ell\bX,t}_{\bx_*(t),t}
	-(\bu \cdot \bnabla\bu)_\delta \big|^{\bx_*(t)+\ell\bX,t}_{\bx_*(t),t}
	= -\frac{1}{\ell} \, \big( \bv \cdot \bnabla_{\bX} \bv \big)_\alpha \big|_{\mathbf{0},t}^{\bX,t}.
	\label{eqA2fB}
	\end{equation}
Derivation of this relation is the most technical part: it uses the properties $G_\delta * \bu = \bu$ and $G_\delta * \bnabla \bu = \bnabla \bu$ with proper changes of variables $\bx \mapsto \bX$ in filter integrals. Verifying also the filtered version of Eq.~(\ref{eqA4b}), one completes the derivation in a similar way. We leave the remaining details as an exercise.


\bibliographystyle{jfm}
\bibliography{biblio}

\end{document}